\documentclass[aps,prd,reprint,groupedaddress]{revtex4-1}
\usepackage[dvipsnames]{xcolor}
\usepackage{graphicx}
\usepackage{hyperref}
\usepackage{url} 
\hypersetup{
    pdfnewwindow=true,      
    colorlinks=true,        
    linkcolor=blue,         
    citecolor=blue,        
    filecolor=blue,         
    urlcolor=blue           
}

\newcommand{\aap}{Astron. Astrophys.}

\newcommand{\apjl}{ApJ Lett.}
\newcommand{\apjs}{ApJS}

\definecolor{pink}{RGB}{255, 20, 147}

\definecolor{carminered}{rgb}{1.0, 0.0, 0.22}

\definecolor{byzantine}{rgb}{0.74, 0.2, 0.64}

\definecolor{amber}{rgb}{1.0, 0.75, 0.0}

\definecolor{amethyst}{rgb}{0.6, 0.4, 0.8}

\definecolor{blue-violet}{rgb}{0.54, 0.17, 0.89}

\definecolor{atomictangerine}{rgb}{1.0, 0.6, 0.4}

\definecolor{comment}{RGB}{166, 38, 164}

\usepackage{amsmath}
\usepackage{amssymb}

\usepackage[acronym]{glossaries}
\setacronymstyle{long-short}
\glsdisablehyper

\newacronym{ligo}{LIGO}{Laser Interferometer Gravitational-wave Observatory}
\newacronym{gw}{GW}{gravitational-wave}
\newacronym{llo}{LLO}{LIGO Livingston Observatory}
\newacronym{er14}{ER14}{LIGO's Engineering Run 14}
\newacronym{o3}{O3}{LIGO's Observing Run 3}
\newacronym{roc}{ROC}{Receiver Operating Characteristic}
\newacronym{snr}{SNR}{signal-to-noise ratio}
\newacronym{far}{FAR}{false-alarm rate}

\begin{document}

\title{Efficient Gravitational-wave Glitch Identification from Environmental Data Through Machine Learning}

\author{Robert E. Colgan$^{1,2}$, K. Rainer Corley$^{3,4}$, Yenson Lau$^{2,5}$, Imre Bartos$^{6}$, John N. Wright$^{2,5}$, Zsuzsa M\'arka$^{4}$, and Szabolcs M\'arka$^{3}$}

\address{$^1$Department of Computer Science, Columbia University in the City of New York, 500 W. 120th St., New York, NY 10027, USA\\
$^2$Data Science Institute, Columbia University in the City of New York, 550 W. 120th St., New York, NY 10027, USA\\
$^3$Department of Physics, Columbia University in the City of New York, 538 W. 120th St., New York, NY 10027, USA\\
$^4$Columbia Astrophysics Laboratory, Columbia University in the City of New York, 538 W. 120th St., New York, NY 10027, USA\\
$^5$Department of Electrical Engineering, Columbia University in the City of New York, 500 W. 120th St., New York, NY 10027, USA\\
$^6$Department of Physics, University of Florida, PO Box 118440, Gainesville, FL 32611-8440, USA}

\begin{abstract}
The LIGO observatories detect gravitational waves through monitoring changes in the detectors' length down to below $10^{-19}$\,$m/\sqrt{Hz}$ variation---a small fraction of the size of the atoms that make up the detector.
To achieve this sensitivity, the detector and its environment need to be closely monitored.
Beyond the gravitational wave data stream, LIGO continuously records hundreds of thousands of channels of environmental and instrumental data in order to monitor for possibly minuscule variations that contribute to the detector noise.
A particularly challenging issue is the appearance in the gravitational wave signal of brief, loud noise artifacts called ``glitches,'' which are environmental or instrumental in origin but can mimic true gravitational waves and therefore hinder sensitivity.
Currently they are primarily identified by analysis of the gravitational wave data stream.
Here we present a machine learning approach that can identify glitches by monitoring \textit{all} environmental and detector data channels, a task that has not previously been pursued due to its scale and the number of degrees of freedom within gravitational-wave detectors.
The presented method is capable of reducing the gravitational-wave detector network's false alarm rate and improving the LIGO instruments, consequently enhancing detection confidence.

\end{abstract}

\maketitle



\section{Introduction}\label{sec:introduction}

Modern interferometric \gls{gw} detectors \cite{2015CQGra..32g4001L, 2015CQGra..32b4001A} are highly complex and sensitive instruments.
Each detector is sensitive not only to gravitational radiation, but also to noise from sources including the physical environment, seismic activity, and complications in the detector itself.
The output data of these detectors is therefore also highly complex.
In addition to the desired signal, the \gls{gw} data stream contains sharp lines in its noise spectrum and non-Gaussian transients, or ``glitches," that are not astrophysical in origin.

\begin{figure*}
    \centering
    \includegraphics[width=0.5\linewidth]{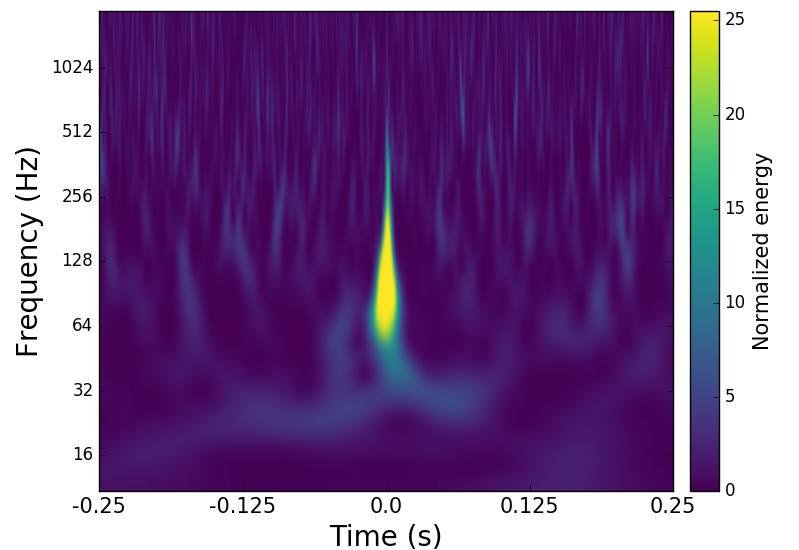}\includegraphics[width=0.5\linewidth]{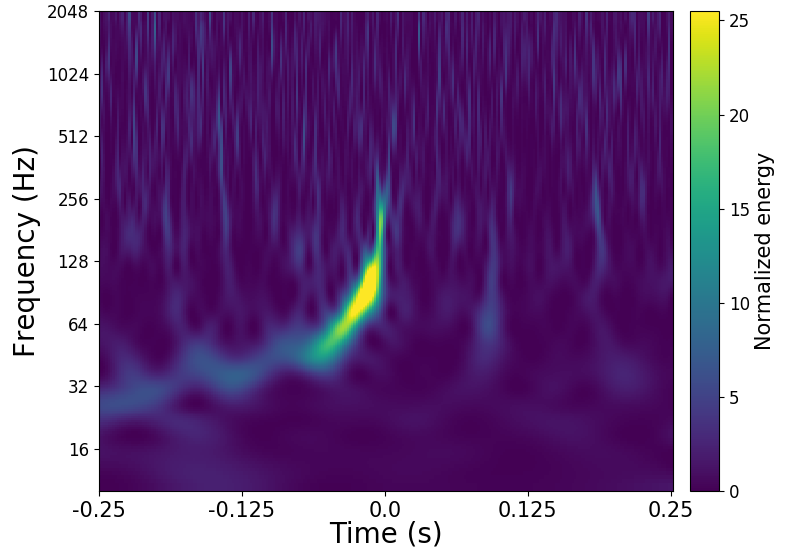}
    \caption{Comparison of the omegagrams \cite{rollins_thesis, 2004CQGra..21S1809C} of a glitch event (left) and \gls{gw} event (right) in LIGO O2 data. Glitches can be mistaken for unmodeled \gls{gw} events and also trigger modeled searches. Images obtained from GravitySpy \cite{2017CQGra..34f4003Z} database.\label{fig:example_glitch}}
\end{figure*}

Instrumental artifacts in the \gls{gw} data stream can be mistaken for short-duration, unmodeled \gls{gw} events, and noisy data can also decrease the confidence in compact binary detections, sometimes by orders of magnitude \cite{2018CQGra..35f5010A}.
We show an example of the similarity between a glitch and a \gls{gw} signal in Fig. \ref{fig:example_glitch} to illustrate the difficulty in searching for \gls{gw} signals with glitches present.
Thus, it is important to identify and flag \gls{gw} data containing glitches.
Flagged instrumental glitches can then be addressed in many ways, from graceful modeling followed by subtraction to the cruder approach of so-called ``vetoes" that unnecessarily waste data.
Understanding the origin of instrumental glitches is also important for diagnosing their causes and improving the quality of the detector and its data.

The primary pipeline currently used to identify and characterize glitches in the Advanced LIGO and Virgo detectors is Omicron \cite{Omicron} (see also \cite{rollins_thesis, 2004CQGra..21S1809C}).
Omicron identifies glitches by searching the \gls{gw} strain data from a single detector for events of excess power.
It characterizes their properties, such as amplitude, duration, and frequency, by comparing the event to a sine-Gaussian waveform.

In addition to the \gls{gw} data stream, each detector records hundreds of thousands of ``channels" of auxiliary data, each channel measuring some aspect of the detector's components or physical environment.
These channels provide important information about the state of the detector that can be useful for diagnosing glitches, but monitoring all of them is a difficult task.

When identifying and flagging potential glitches, it is important to ensure that the event is indeed a glitch of instrumental origin, rather than an unmodeled \gls{gw} event.
By considering only LIGO's auxiliary channels, rather than the \gls{gw} data stream, we can be more confident that glitches we flag are indeed not from gravitational events, since the auxiliary channels are generally not sensitive to \glspl{gw} \cite{2016CQGra..33m4001A,2010JPhCS.243a2005I}.
An additional benefit of studying glitches with auxiliary channels is that correlations with specific channels can help identify the source of detector issues.

Current work to veto glitch segments includes UsedPercentageVeto \cite{2010JPhCS.243a2005I}, HierarchicalVeto \cite{2011CQGra..28w5005S}, Bilinear Coupling Veto \cite{2014PhRvD..89l2001A}, and iDQ \cite{2013CQGra..30o5010E, 2013PhRvD..88f2003B,essick_thesis}.
Cavaglia et al. \cite{CiCP-25-963} use a dual approach of random forests combined with genetic programming in order to identify instrumental artifacts from a subset of auxiliary channels.
Additionally, there is work to study data quality by correlating auxiliary channels with the \gls{gw} detector's astronomical range \cite{2018CQGra..35v5002W}.
Our method is complementary to these approaches; we consider \textit{all} auxiliary channels and use them to identify segments containing glitches, both to flag these segments and to identify detector issues.

Machine learning techniques have proved to be powerful tools in analyzing complex problems by learning from large example datasets.
They have been applied in \gls{gw} science from as early as 2006~\cite{Lightman_2006} to the study of glitches \cite{2013CQGra..30o5010E,2013PhRvD..88f2003B,2018arXiv181205225C,2015CQGra..32u5012P,2017CQGra..34c4002P,2017CQGra..34f4003Z,2017PhRvD..95j4059M,2018CQGra..35i5016R,2017arXiv171107468G,2018PhRvD..97j1501G} and other problems, such as real-time signal detection \cite{2018PhRvD..97d4039G}, signal characterization \cite{2017PhRvD..96j4015K,2017CQGra..34i4003V,2018PhRvL.120n1103G}, and parameter estimation \cite{2018PhLB..778...64G}.

Classification is a fundamental problem in machine learning in which a machine learning model is trained to consider a set of measured characteristics, or ``features," of data samples belonging to one of at least two categories.
By providing the model with a set of samples whose categories are known, we can train it to predict the category of samples whose true category is unknown.

We pose the problem of detecting whether a glitch is occurring at a given time based on LIGO auxiliary channels as a simple two-class classification problem (as in previous work \cite{2013PhRvD..88f2003B,CiCP-25-963}) and apply a well-understood, efficient, and commonly used machine learning method to this problem, with promising results.

We train a classification model to predict whether a glitch is occurring at a given time using features derived from the \gls{gw} detector's auxiliary channels at that time.
Because the data on which the classification is performed are derived only from the auxiliary channels, our method is able to provide corroboration of the presence or absence of glitches without using strain data, independently of existing methods that analyze the strain.

Below, we describe a variant of the method, called Elastic-net based Machine-learning for Understanding (EMU method or method thereafter), we use to identify glitches in Sec. \ref{sec:methods}; we show the results of testing this method on recent LIGO data in Sec. \ref{sec:results}; and we discuss the results in Sec. \ref{sec:conclusion}.


\section{Methods}\label{sec:methods}
To predict the presence of a glitch in a \gls{gw} data stream using auxiliary information, we need a method that extracts useful information from the auxiliary channels and uses this information to make a decision.

In considering all of the auxiliary channels, we must generate and process a large, high-dimensional dataset for classification.
Linear models are simple and effective on this type of large dataset, and they are straightforward to train.
We use logistic regression, a standard linear model for binary classification.
We choose as the feature set a group of representative statistics for each auxiliary channel to capture properties of the channel's behavior in the vicinity of a glitch (or absence of one).

If we hope to also diagnose detector issues, it is important that we can interpret the output of the algorithm.
Logistic regression provides a simple method for this: during training, each feature is given a weight; higher-magnitude weights indicate features that are more relevant in deciding whether a glitch is present in a sample.
A model with fewer large weights is intuitively more easily interpretable.
To encourage this property, we use elastic net regularization to penalize the weights such that only the most relevant features are selected by the model and those corresponding to uninformative features become 0.
This also allows us to train with smaller datasets than would otherwise be required if we did not impose sparse regularization.

Below, we describe the preliminary step of identifying irrelevant data in Sec. \ref{sec:data_reduction}, extraction of features from the remaining data in Sec. \ref{sec:features}, pre-conditioning of the data in Sec. \ref{sec:conditioning}, the machine learning model we employ in Sec. \ref{sec:model}, selection of model hyperparameters in Sec. \ref{sec:hyperparameters}, and an analysis of how much training data is sufficient in Sec. \ref{sec:train_data_amt}.

\subsection{Preliminary data reduction}\label{sec:data_reduction}
There are approximately 250,000 auxiliary channels in each LIGO detector. 
Many of these channels are constant or always change in a consistent pattern (e.g., tracking the time or counting CPU cycles), and can be safely ignored.
To reduce the amount of data that must be processed, we remove these channels that are uninformative for glitches from the analysis.

Auxiliary channel time-series data is encoded in a custom format \cite{gwf-format} and stored in files each containing 64 consecutive seconds of data for each channel; we refer to these files as ``raw frame files."
To identify uninformative channels, we choose a few raw frame files from the training period of each analysis and compare the channels across those frames.
For each channel in each of the selected frames, we subtract the channel's first raw value from the following values in that frame, and compare the resulting time series to the corresponding one from each of the other selected frames.
If all are identical, the channel is ignored for the rest of the analysis.
After this procedure, approximately 40,000 channels remain for our further analysis.

Some auxiliary channels are directly coupled with the \gls{gw} strain channel, or are contaminated in other ways by the \gls{gw} signal.
Many of these channels have been identified in internal LIGO ``channel safety" studies \cite{2016CQGra..33m4001A,2010JPhCS.243a2005I}.
We removed all such known channels.
Additionally, after a preliminary training run and discussions with experts, we removed additional channels that may be contaminated by the strain but had not been considered in safety studies.

\subsection{Feature extraction}\label{sec:features}
We use the Omicron \cite{Omicron} event trigger generator to identify glitch times for training and testing our model.
(Once trained, our model is fully independent of Omicron and the strain data; it considers only parameters computed from auxiliary channels, as described below.)
Omicron analyzes low-latency strain data to find events of excess power and reports parameters of these glitches, including start time, peak time, and duration.

For our training, we gather points in time (i) drawn from the peak times of glitches (``glitchy" times), and (ii) drawn from stretches of at least four seconds with no recorded glitches (``glitch-free" times).
For the glitch-free samples, we select times such that no part of any glitch (accounting for its full duration) falls within two seconds before or after the sample time.
(We note that this does not mean we veto four seconds of data for each potential glitch our method flags; it simply means that during the training process we require the time segment around our glitch-free training examples to be sufficiently clean.)

For each glitchy or glitch-free point in time, we generate an array of ten statistical quantities (described in the following paragraphs) for each channel to characterize the channel's behavior around that time.
These quantities become the features for our analysis.

\begin{figure}
    \centering
    \includegraphics[width=\linewidth]{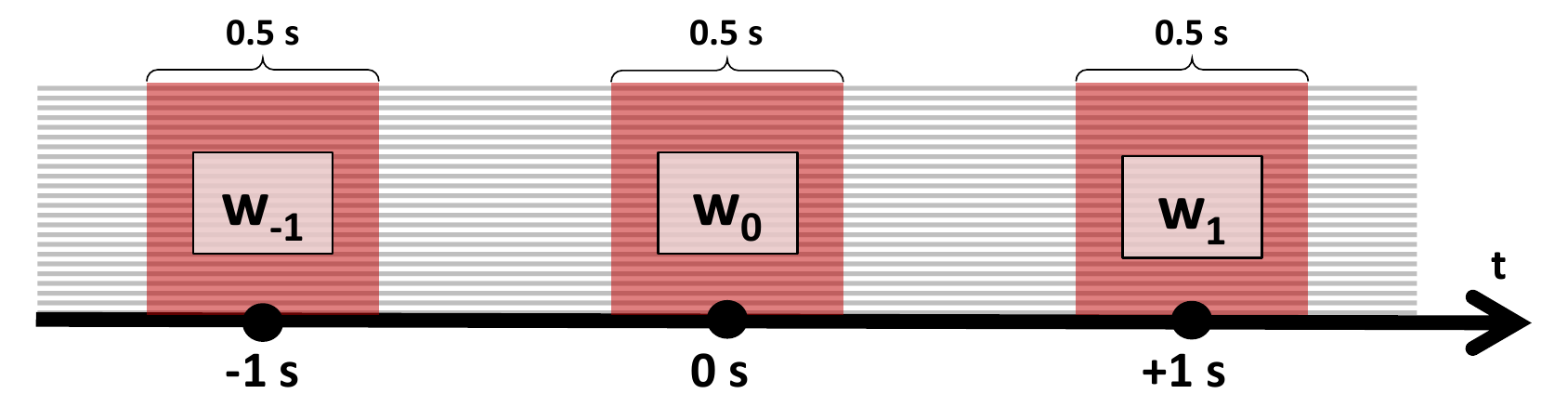}
    \caption{Illustration of the time intervals considered in statistical feature array.\label{fig:time_intervals}}
\end{figure}

Let us denote a given glitch peak time or glitch-free sample time as $t_0$ and the times one second before and one second after as $t_{-1}$ and $t_1$ respectively.
We consider three time windows of 0.5 seconds duration centered at $t_{-1}$, $t_0$, and $t_1$, denoted $w_{-1}$, $w_0$, and $w_1$.
This is illustrated in Fig. \ref{fig:time_intervals}.

For a given sample time $t_0$, for each channel we construct the following ten-dimensional vector based on the three time windows:

\newcommand{\spc}{\;\;\;}
\begin{subequations}
    \begin{align}
        \mathbf{v} = \Bigl(\ &\mu_{-1},\spc \mu_{0},\spc  \mu_{1},\spc \sigma_{-1},\spc \sigma_{0},\spc \sigma_{1}, \label{eqn:feature_a}
        \\
        &\mu_{1} - \mu_{-1}, \spc\spc\spc\spc\spc \sigma_{1} - \sigma_{-1}, \label{eqn:feature_b}
        \\
        &\mu_{0} - \frac{\mu_{1}+\mu_{-1}}{2}, \spc\spc \sigma_{0} - \frac{\sigma_{1}+\sigma_{-1}}{2}\ \Bigr),\label{eqn:feature_c}
    \end{align}
\end{subequations}
where $\mu_i$ and $\sigma_i$ are the mean and standard deviation, respectively, of the channel's output over the time window $w_i$, with $i \in \{-1, 0, 1\}$.
(We note that this does not mean we veto 2.5 seconds of data for each potential glitch our method flags; it simply means that for a given channel and sample time, we consider the surrounding time period so we can identify deviations from the channel's behavior in its local environment, as implied in the definition of the features.)
The time and duration of the transient can be extracted as the feature definition is known.

In contrast to \cite{2013PhRvD..88f2003B}, which considers auxiliary data in a 100 millisecond window around glitch transients, our method considers data from a span of 2.5 seconds, enabling longer timescale couplings to be addressed.
We also consider many more channels in our analysis.

Each feature was chosen with the intent of capturing certain properties of a channel's behavior in the vicinity of a glitch or the absence of one.
The features in \ref{eqn:feature_a} were chosen to capture the mean and standard deviation of the channel's raw value shortly before, during, and shortly after $t_0$.
Those in \ref{eqn:feature_b} were chosen to identify step changes occurring near $t_0$.
Those in \ref{eqn:feature_c} were chosen to identify short, temporary changes occurring near $t_0$.

It should be noted that these features were chosen ad-hoc based on intuition of what properties of channels' behavior might be informative.
Compared to extracting features from nearby auxiliary transients in various wavelet domains (see e.g.~\cite{2013PhRvD..88f2003B,2018arXiv181205225C}), the method presented here is simpler and requires fewer computational resources, especially when considering the significantly increased dimensionality of the problem addressed here.
We believe the simplicity of these features is an advantage of our method, but it is otherwise essentially agnostic to the features chosen here; more descriptive features could potentially improve its performance.
We leave this exploration to future work.

We construct the vector $\mathbf{v}$ for each of the approximately 40,000 channels in consideration, resulting in approximately 400,000 features for each glitchy or glitch-free point in time.

\subsection{Data pre-conditioning}\label{sec:conditioning}
Most machine learning techniques assume that each feature is on approximately the same scale; otherwise features whose raw values are large in magnitude would dominate the others.
A standard normalization procedure is to replace raw values with their standard score (i.e., the number of standard deviations away from the training mean that the raw value falls), so each feature has zero mean and unit standard deviation over the training set \cite{Bishop, Hastie}.
For each analysis under consideration, we compute the mean and standard deviation over the training set; then for every point in the training, validation, and test sets, we subtract the mean and divide by the standard deviation of that feature in the training set.

Occasionally, the raw channel data contains missing or invalid values, resulting in invalid entries in our feature matrix.
When this occurs, we simply replace the entry with the mean of the valid entries for that feature in the training set prior to performing the normalization described above.

\subsection{Glitch classification via logistic regression}\label{sec:model}
We formulate the problem of identifying glitches as a basic statistical classification problem, where instances in which a glitch is present are classified as 1 (``glitchy") and instances where no glitch is present are classified as 0 (``glitch-free").

We use logistic regression with elastic net regularization to perform this classification using the features derived from auxiliary channel data described in Sec. \ref{sec:features}.

Logistic regression is a well-established linear classification method in statistics and machine learning \cite{Bishop, Hastie}.
It is related to classical linear regression, but rather than predicting a continuous unbounded variable, a logistic function is applied to the output to restrict it between 0 and 1.

Given a set of $n$ training data points in $p$ dimensions (i.e., each data point has $p$ features) and $n$ corresponding binary labels (the ground truth), a logistic regression model is trained by iteratively minimizing the residual error between the predicted class probability of the training data and ground truth.
The trained model consists of a set of $p$ coefficients (or ``weights") $\mathbf{w}$ and a bias term $b$; the dot product of these coefficients and a test data point plus the bias term is passed through a logistic function to obtain an estimate of the probability that the point should be classified 0 or 1.

Let $\sigma(\cdot)$ denote the logistic function:
\begin{equation}\label{eqn:logistic_sigma}
    \sigma(a) = \frac{1}{1+\exp(-a)}
\end{equation}
Then the probability estimated by the model that a test data point $\mathbf{x}$ belongs to the class 1 is:
\begin{equation}\label{eqn:logistic_prob}
    P(\mathbf{x}=1) = \sigma(\mathbf{w}^\mathrm{T} \mathbf{x} + b)
\end{equation}
This value may be thresholded to produce a binary output.
The threshold is commonly 0.5 but may be chosen as desired to adjust the ratio of false positives and false negatives.

During training, a measure of the error between the known ground truth label $y_i \in \{0, 1\}$ and the current model's prediction $\sigma(\mathbf{w}^\mathrm{T} \mathbf{x}_i + b)$ for a training point $\mathbf{x}_i \in \mathbb{R}^{p}$ can be quantified:
\begin{equation}\label{eqn:lr_err}
\begin{split}
    E_{\mathbf w, b}(y_i,\mathbf{x}_i) = -\Bigl(&y_i \log(\sigma(\mathbf{w}^\mathrm{T} \mathbf{x}_i + b)) +\\
    &(1 - y_i) \log(1 - \sigma(\mathbf{w}^\mathrm{T} \mathbf{x}_i + b))\Bigr)
\end{split}
\end{equation}
Known as the cross-entropy error, this is a convex function that can be minimized over $\mathbf{w}$ and $b$ by gradient descent or other iterative methods \cite{Bishop, Hastie}.

Various regularization terms may be applied to the coefficients and added to the residual error during training as a penalty, to reduce overfitting and induce desired properties in the trained model \cite{Bishop, Hastie}.
Let $R(\mathbf{w})$ be some regularization function for the coefficients $\mathbf{w}$. The combined cost (or ``loss'') function that is iteratively minimized during training is given by:
\begin{equation}\label{eqn:lr_loss}
    L(\mathbf{w},b) = \frac{1}{n} \sum^n_{i=1} E_{\mathbf w, b}(y_i,\mathbf{x}_i) + \alpha R(\mathbf{w}),
\end{equation}
where $\alpha$ is a hyperparameter controlling the overall regularization strength relative to the error term $E$.

Common choices of regularization function with logistic regression are the L2 norm (also known as ridge regression or Tikhonov regularization) and the L1 norm (the same penalty used in the LASSO \cite{LASSO}).
In addition to mitigating overfitting by penalizing the overall magnitude of the coefficient vector, the L1 norm also induces sparsity in the coefficients (i.e., many of them will be zero).
This is often desirable for scalability and interpretability when the dimension $p$ of the input data is high, as is the case with our dataset.
After training, the nonzero coefficients suggest which of the input features are most important in determining the classification result \cite{Hastie}.

\begin{figure*}
    \centering
    \includegraphics[width=\textwidth]{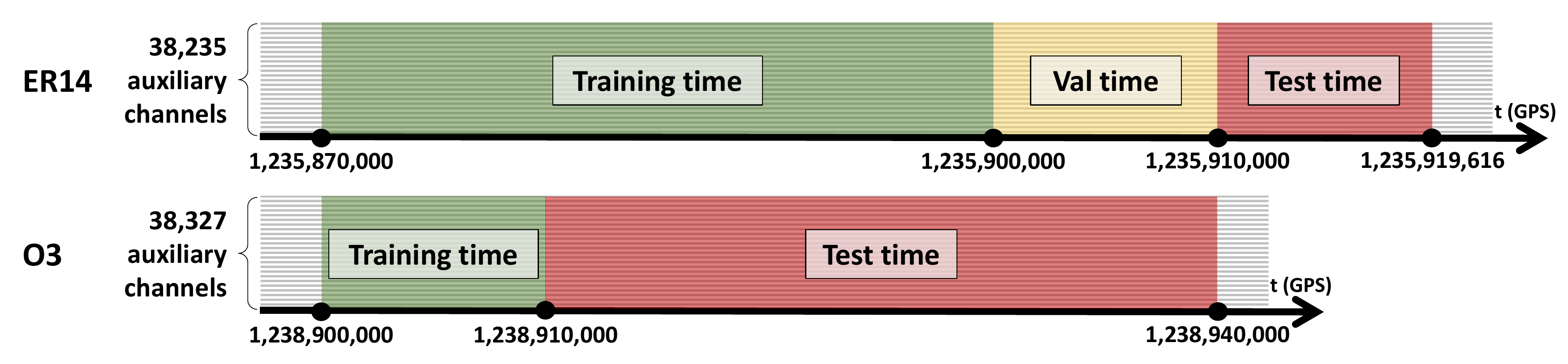}
    \caption{Time periods used to create training, validation, and test datasets. The \gls{er14} analysis includes a validation dataset for tuning hyperparameters, which were fixed for all subsequent tests, so there is no validation period for the \gls{o3} analysis.
    \label{fig:er14_o3_train_val_test_times}}
\end{figure*}

The elastic net \cite{10.2307/3647580} is a weighted sum of the L1 and L2 norms:
\begin{equation}\label{eqn:elastic}
    R(\mathbf{w}) = \frac{\lambda}{2} \sum^p_{j=1} w_j^2 + (1 - \lambda) \sum^p_{j=1} |w_j| \end{equation}
where $\lambda$ is a hyperparameter controlling the relative strength of the L1 and L2 regularization.
The elastic net also induces sparsity, but less strongly than L1.
It also has the desirable advantage over L1 of being more likely to select correlated features together rather than arbitrarily choosing only one of them \cite{10.2307/3647580}.
In our application, if features from many channels are correlated, it may be useful for diagnostic purposes to consider all of them rather than only one.

\subsection{Hyperparameter optimization}\label{sec:hyperparameters}
Elastic net logistic regression has two hyperparameters that must be tuned to achieve the best result: overall regularization strength $\alpha$, in Eq. \ref{eqn:lr_loss}, and the ratio $\lambda$ of the strength of L1 and L2 regularization, in Eq. \ref{eqn:elastic}.

We performed a grid search across a range of both parameters using data from \gls{llo} during \gls{er14} and evaluated the results on a held-out validation dataset drawn from a separate period of time (see Fig. \ref{fig:er14_o3_train_val_test_times}).
Using a validation dataset separate from both the training set and the test set on which we report our results allows us to choose hyperparameters that will generalize well to unseen data without overfitting to our training or test data \cite{Bishop, Hastie}.

We trained models over a grid of $\alpha$ and $\lambda$ values on a dataset drawn from 30,000 seconds during a lock segment on March 6, 2019 (GPS time 1,235,870,000 to 1,235,900,000).
To create the training dataset, we randomly sampled 7,500 glitch-free points in time (as described in Sec. \ref{sec:features}) and 7,500 of the 30,141 Omicron glitches during that period.
This subsampling was performed to allow the dataset to fit in available memory.
For both datasets, we also ignored any samples falling too close to the beginning or end of a 64-second raw frame file.
After preliminary data reduction as described in Sec. \ref{sec:data_reduction}, the number of channels considered was 38,235.
We then generated the 382,350 statistical features for each point, as described in Sec. \ref{sec:features}.
We then trained an elastic net logistic regression model independently for each $\alpha, \lambda$ pair on this training set.
Training was performed using the Scikit-learn package \cite{scikit-learn}.

We evaluated each trained model on a validation dataset drawn from the 10,000 seconds immediately following the training period (GPS time 1,235,910,000 to 1,235,920,000).
Fig. \ref{fig:er14_o3_train_val_test_times} illustrates the times these ER14 datasets are drawn from.
The validation dataset was created similarly to the training set, by sampling 2,500 glitch-free points and 2,500 of the 7,222 Omicron glitches during that period.
We chose the $\alpha, \lambda$ pair that gave the best accuracy on the validation dataset and fixed the values of those hyperparameters based on these results for all further training.

\begin{figure}
    \centering
    \includegraphics[width=\linewidth]{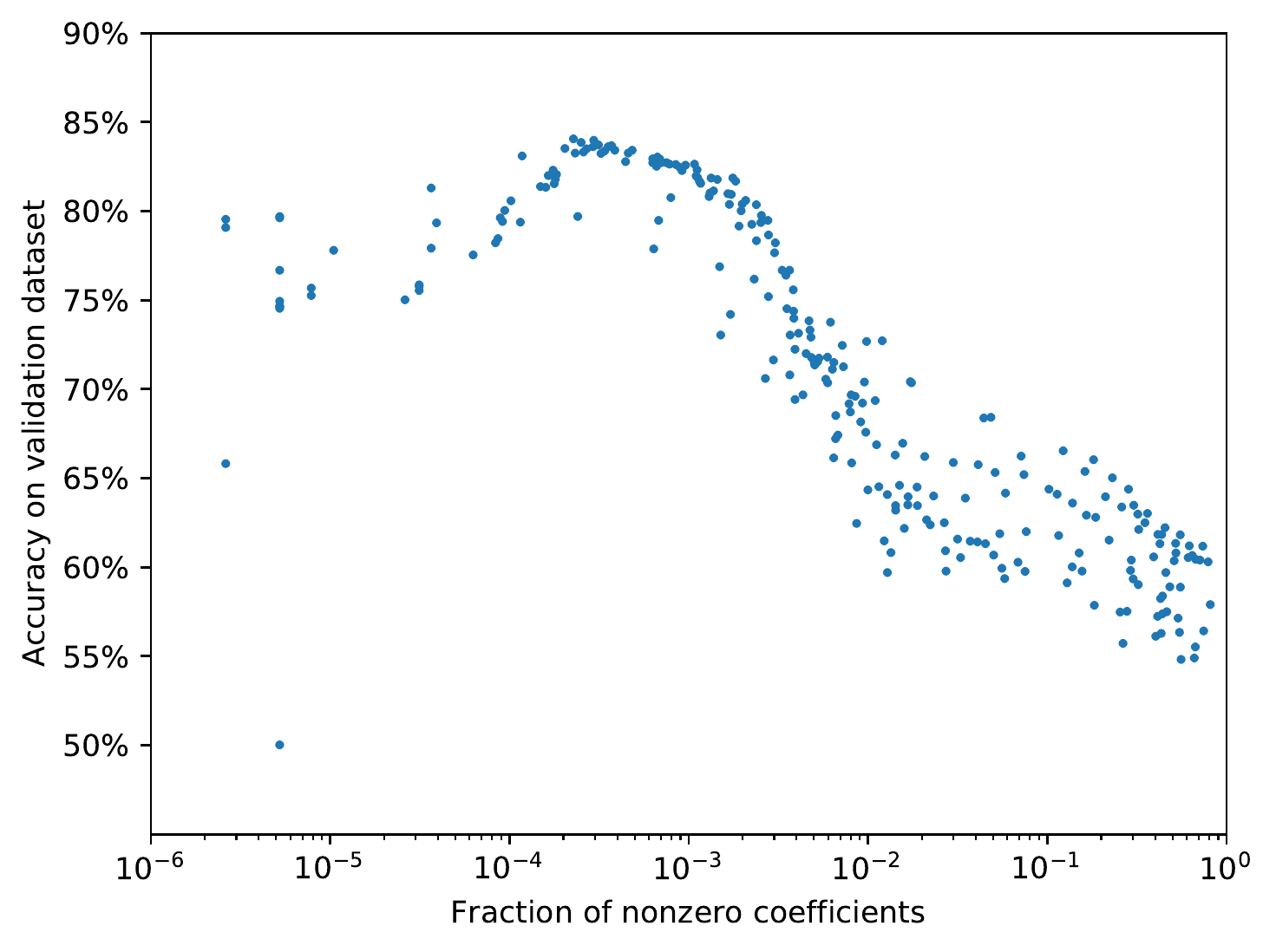}
    \caption{\gls{er14} validation accuracy vs. fraction of nonzero coefficients in the trained model. The models have approximately $p=400,000$ coefficients, so a model with $10^{-3}p$ nonzeros contains only 400 nonzero coefficients. 
    \label{fig:nnz_acc_scatter}}
\end{figure}

Varying the $\alpha$ and $\lambda$ hyperparameters effectively tunes the strength with which the model's coefficients $w_j$ are driven to zero by regularization during training.
After training with many pairs of these parameters, we can evaluate the relationship between the number of nonzero coefficients and the model's predictive accuracy on the validation data.
One would expect that a model with too many zero coefficients would not be able to consider enough features to make accurate predictions, while a model with too many nonzero coefficients would become overfit to the training data and not generalize well to separate data.
Fig. \ref{fig:nnz_acc_scatter} demonstrates that this is the case with our data.

The model that achieves the highest accuracy on our validation dataset, at 84.1\%, contains only 87 (0.02\%) nonzero coefficients.
These coefficients correspond to features from only 56 distinct channels, indicating specific channels of potential detector issues at training time.
None of these 56 channels are known to be coupled with or contaminated by the \gls{gw} strain.
This demonstrates that a small subset of the features and channels are sufficient to consider for glitch classification.
It also demonstrates that elastic net regularization is beneficial not only to interpretability but also to prediction accuracy.

\subsection{Amount of training data}\label{sec:train_data_amt}

\begin{figure}
    \centering
    \includegraphics[width=\linewidth]{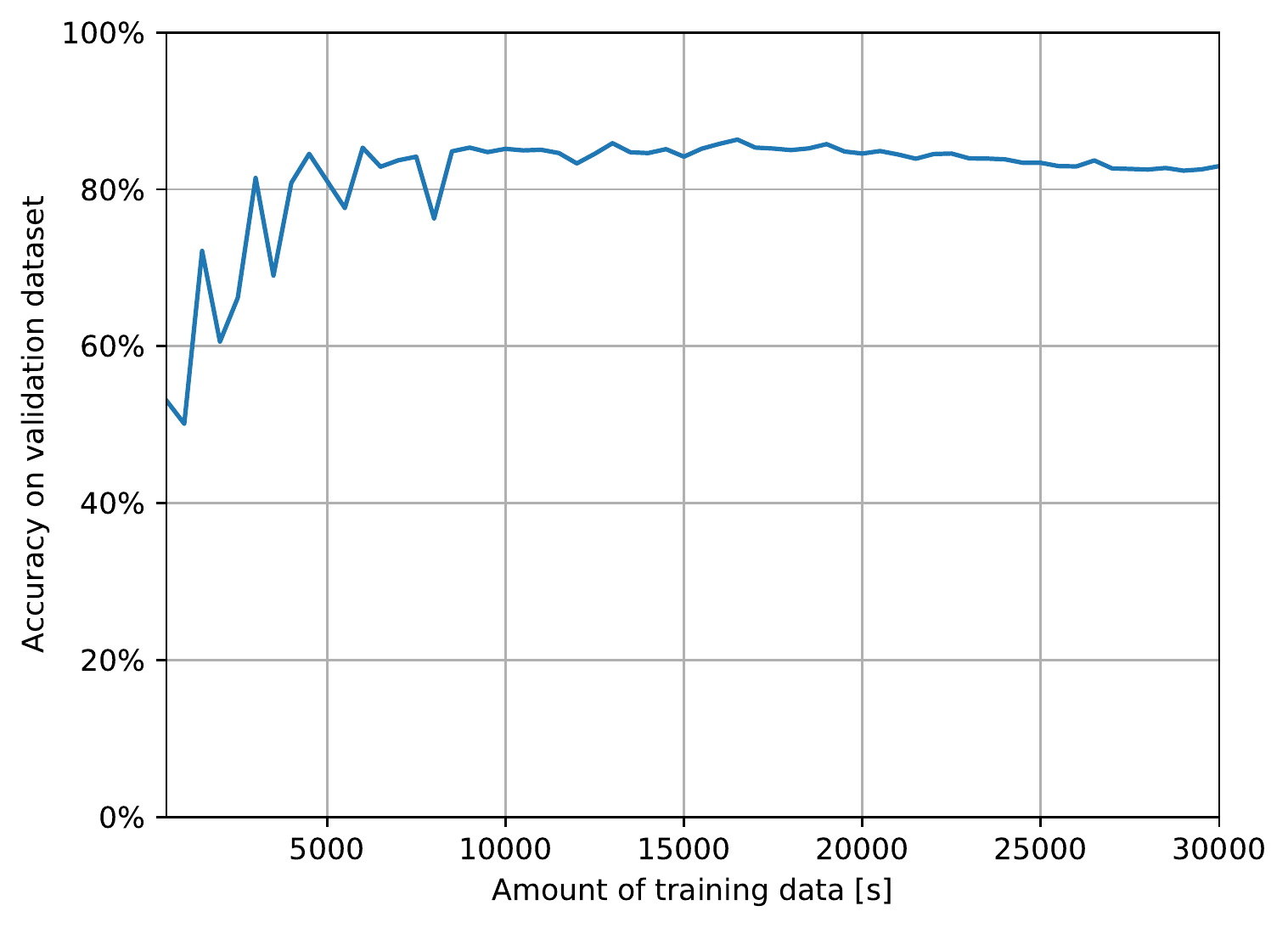}
    \caption{Classifier's accuracy vs. length of training period \label{fig:data_amt_acc_er14}}
\end{figure}

We also investigated how much training data was sufficient for good performance, using the \gls{er14} training and validation data.
For this experiment, we trained a new classifier using data drawn from time periods of varying lengths.
Each of these time periods was a subset of the original training data period, ending at the end of the original training period and starting between 500 and 30,000 seconds earlier.
We show the accuracy of the classifier on the validation dataset (described in \ref{sec:hyperparameters}) for these varying training lengths in Fig. \ref{fig:data_amt_acc_er14}.
The results indicate that 10,000 seconds is a sufficient length of time from which to draw training data. 

\section{Results}\label{sec:results}

We use the EMU method described above to classify glitchy and glitch-free times for several recent segments of data from \gls{llo}.
We show two test cases from different, recent runs: one from a lock segment during \gls{er14}, and a second from a lock segment during \gls{o3}.

The classifier model for each analysis was trained independently using data drawn from a time period close to but not overlapping with the time period from which the test dataset for that analysis was drawn.
We do not perform hyperparameter optimization as we did for \gls{er14} in Sec. \ref{sec:hyperparameters} again for \gls{o3}, because the procedure is computationally intensive and the similar nature of the data means the optimal hyperparameters would likely not be significantly different.

For each analysis, we create a test dataset drawn from a period of time separate from the training dataset (and, in the case of \gls{er14}, separate from the validation dataset as well), as illustrated in Fig. \ref{fig:er14_o3_train_val_test_times}.
We standardize the test dataset according to the means and standard deviations of the features in the training dataset.
We then pass the test dataset through the classifier without labels and compare the predictions to the known ground truth for each point.
For each incorrect classification, we can specify whether the result is a false positive (a glitch-free time classified as glitchy) or a false negative (a glitchy time classified as glitch-free).

\begin{figure}
    \centering
    \includegraphics[width=\linewidth]{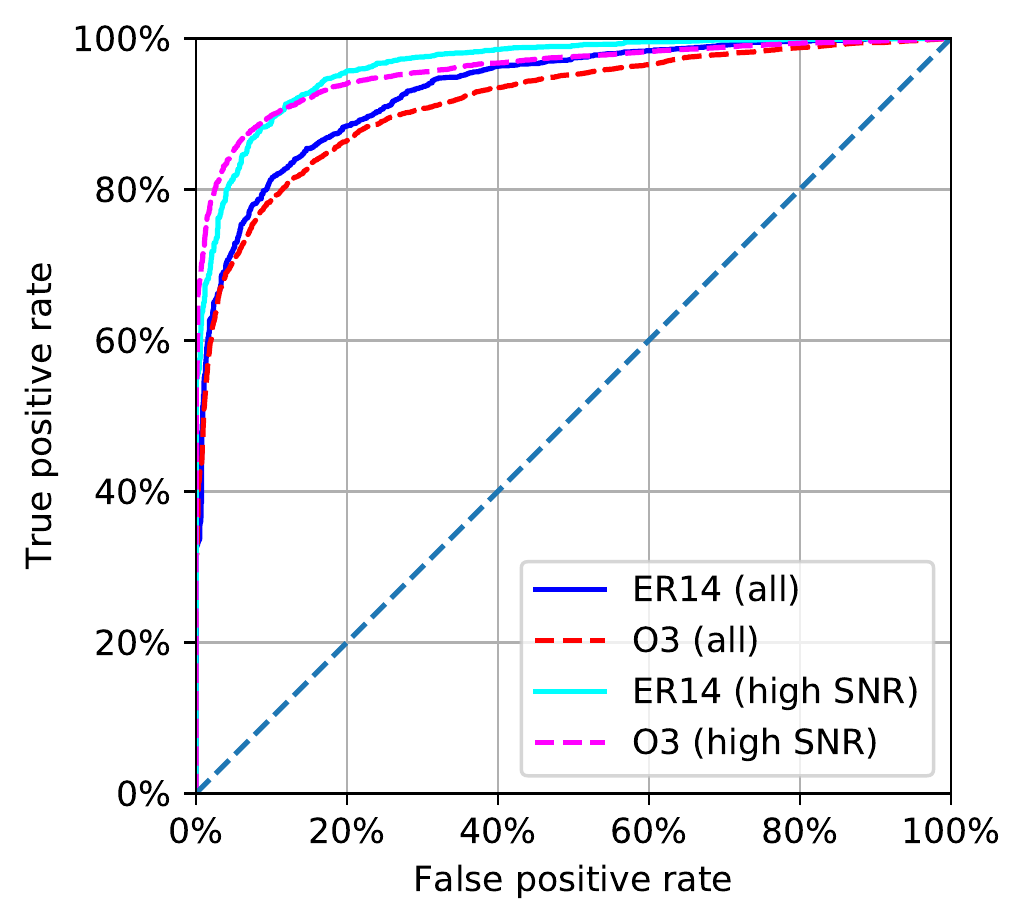}
    \caption{Overall \gls{roc} curves for \gls{er14} and \gls{o3} test segments, and corresponding \gls{roc} curves for only high-SNR glitches.\label{fig:roc_ER14_O3}}
\end{figure}

Note that the actual output of the classifier is the predicted probability of a glitch, which ranges between 0 and 1.
Prior to calculating all reported accuracies, we threshold this value at 0.5 so values at or above are considered predictions of glitches and values below are considered predictions of the absence of a glitch.
We can adjust this threshold to control the ratio of true positives and false positives as necessary for different applications, which might call for a lower false negative rate at the expense of a higher false positive rate or vice versa.
The trade-off is illustrated for both analyses in the \gls{roc} curve in Fig. \ref{fig:roc_ER14_O3}.

At the default decision threshold of 0.5, the accuracy on the \gls{er14} test dataset is 83.8\%, with a true positive rate of 73.0\% and a true negative rate of 94.6\%.
The accuracy on the \gls{o3} test dataset is 79.9\%, with a true positive rate of 62.1\% and a true negative rate of 97.7\%.
We also show the overall accuracy, true positive rate, and true negative rate over time during the test periods for each analysis in Figs. \ref{fig:acc_tp_tn_er14} and \ref{fig:acc_tp_tn_o3}.

If we restrict our analysis in training and testing to glitches with a \gls{snr} at or above 6, we achieve an overall accuracy of 88.2\% for \gls{er14} and 90.3\% for \gls{o3}, with true positive rates of 80.8\% and 86.7\% and true negative rates of 95.4\% and 93.8\% respectively.
(The minimum \gls{snr} reported by Omicron is 5; roughly 30\% of glitches have an \gls{snr} at or above 6.)
The corresponding \gls{roc} curves are displayed on Fig. \ref{fig:roc_ER14_O3}.

\subsection{\gls{er14}}\label{sec:results-er14}

For the \gls{er14} test analysis, we use the trained classifier model that performed best on the \gls{er14} validation dataset used for hyperparameter optimization, as described in Sec. \ref{sec:hyperparameters}.
The training dataset is therefore the same as described there.
Recall that to create this dataset we sampled 7,500 glitch-free points in time (as described in Sec. \ref{sec:features}) and 7,500 of the 30,141 Omicron glitches during the period between GPS times 1,235,870,000 and 1,235,900,000.
After preliminary data reduction as described in Sec. \ref{sec:data_reduction}, the number of channels considered was 38,235, so the training and test datasets have 382,350 features.

The nonzero coefficients of a trained classifier indicate which features the classifier considers when making decisions.
As discussed in Sec. \ref{sec:hyperparameters}, the classifier that performed best on the \gls{er14} validation data had 87 nonzero coefficients corresponding to features from 56 different channels.

The test data is drawn from 9,616 seconds between GPS times 1,235,910,000 and 1,235,919,616.
We sampled 2,500 glitch-free points in time and 2,500 of the 6,479 Omicron glitches during that period, chosen at random.
As with the training, we also ignored any samples falling too close to the beginning or end of a 64-second raw frame file.
The classifier achieves an accuracy of 83.8\% on this test dataset, with a true positive rate of 73.0\% and a true negative rate of 94.6\%.

\begin{figure}
    \centering
    \includegraphics[width=\linewidth]{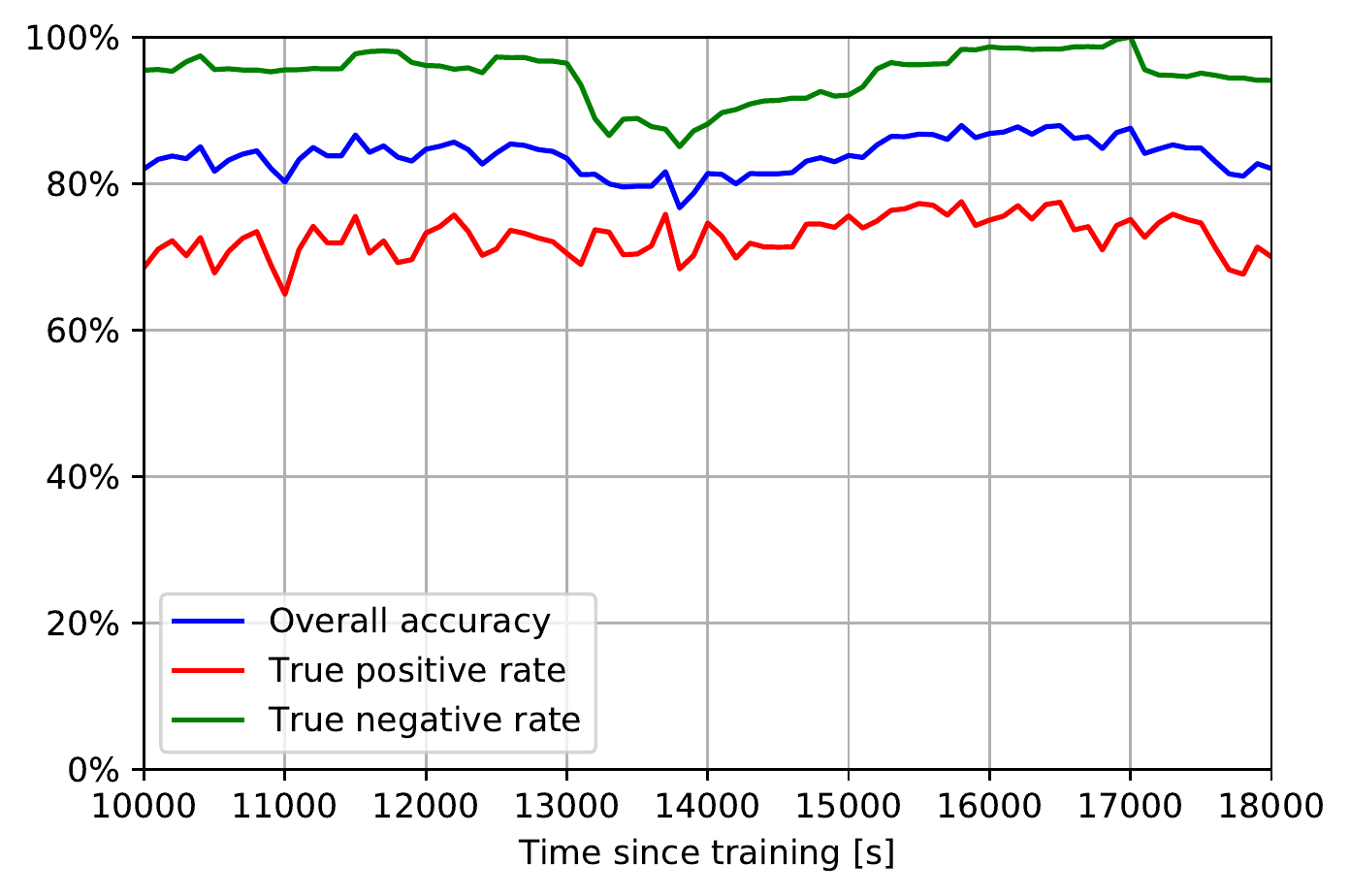}
    \caption{Accuracy, true positive rate, and true negative rate over time during the \gls{er14} test period. For stability, these quantities are computed over rolling windows of length 2,000~s beginning at the indicated time. The x-axis begins at 10,000~s because the 0 to 10,000~s period is used for validation (see Fig.~\ref{fig:er14_o3_train_val_test_times}). 
    \label{fig:acc_tp_tn_er14}}
\end{figure}

The classifier's accuracy, true positive rate, and true negative rate over time in the test period is shown in Fig. \ref{fig:acc_tp_tn_er14}.
This result indicates that the performance is relatively consistent over time, but may be affected by transient changes in the state of the detector that were not seen during training.
\begin{figure*}
    \centering
    \includegraphics[width=\linewidth]{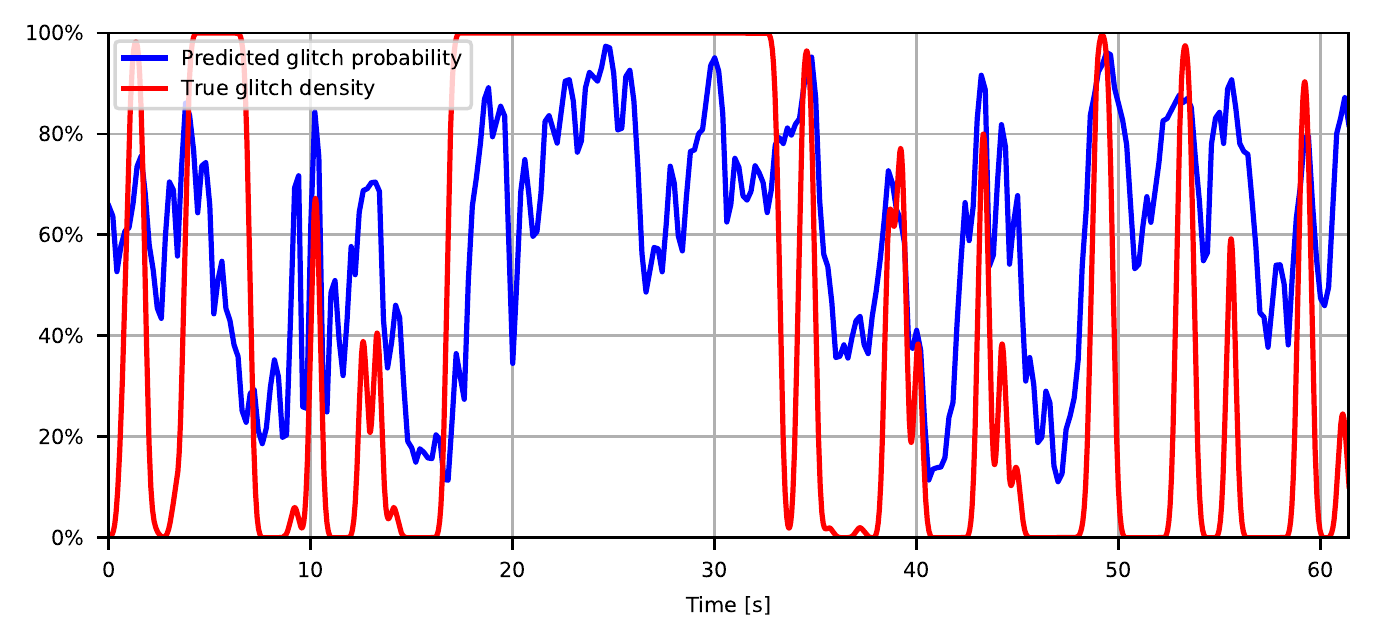}
    \caption{Visualization of estimated glitch probability (blue) and true glitch density (red) over time during a 64-second segment from early \gls{o3}. (Glitch density was calculated by checking whether Omicron reported a glitch of any \gls{snr} during each sample time, then smoothing the resulting binary vector by convolving it with a Gaussian with $\sigma=0.2$ s.)
    \label{fig:lr_prob_over_time}}
\end{figure*}

\subsection{\gls{o3}}\label{sec:results-o3}

For the \gls{o3} analysis, we trained a new classifier on a training dataset drawn from 10,000 seconds during April 10, 2019 (GPS times 1,238,900,000 to 1,238,910,000).
We used a smaller amount of time for the training period for this analysis because the results shown in Fig. \ref{fig:data_amt_acc_er14} indicate that 10,000 seconds of training data is sufficient for good performance.
Similarly to the \gls{er14} training dataset, we sampled 2,500 glitch-free points in time and 2,500 of the 8,098 Omicron glitches during that period, chosen at random, and we ignored any samples falling too close to the beginning or end of a 64-second raw frame file.
After preliminary data reduction as described in Sec. \ref{sec:data_reduction}, the number of channels considered is 38,327, so the training and test datasets have 383,270 features.

After training, the \gls{o3} classifier had 55 nonzero coefficients corresponding to features from 46 distinct channels.

The test data is drawn from 30,000 seconds between GPS times 1,238,910,000 and 1,238,940,000.
We sampled 7,500 glitch-free points in time and 7,500 of the 24,243 Omicron glitches during that period, chosen at random, and we ignored any samples falling too close to the beginning or end of a 64-second raw frame file.
The classifier achieves an accuracy of 79.9\% on this test dataset, with a true positive rate of 62.1\% and a true negative rate of 97.7\%.

\begin{figure}
    \centering
    \includegraphics[width=\linewidth]{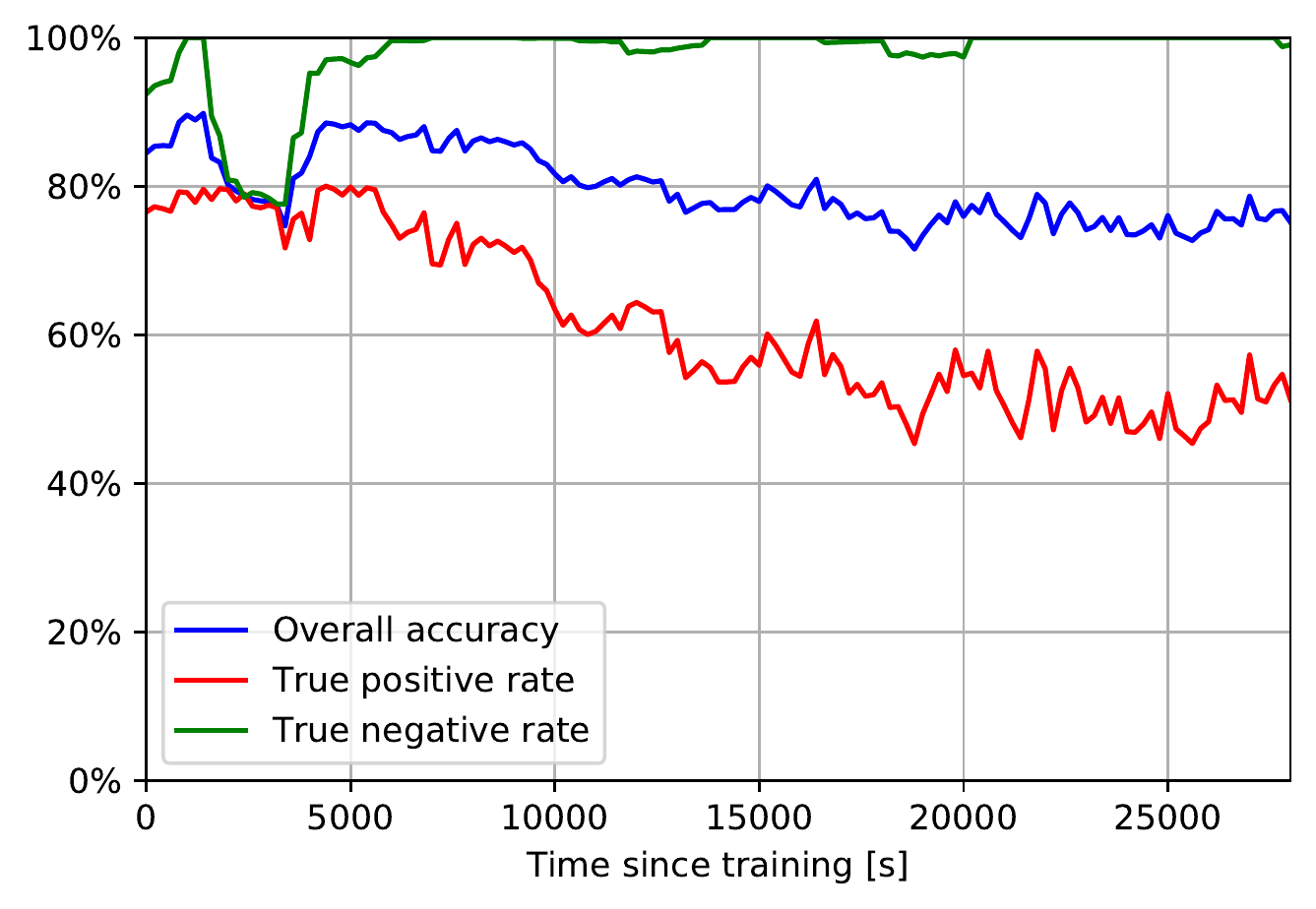}
    \caption{Accuracy, true positive rate, and true negative rate over time during the \gls{o3} test period. For stability, these quantities are computed over rolling windows of length 2,000~s beginning at the indicated time. Note that the scale of the x-axis is different from Fig.~\ref{fig:acc_tp_tn_er14}.
    \label{fig:acc_tp_tn_o3}}
\end{figure}

The accuracy, true positive rate, and true negative rate over time during the test period is shown in Fig. \ref{fig:acc_tp_tn_o3}.
As with \gls{er14}, this result indicates that the performance is relatively consistent over time, but there is a visible dip in true negative rate and corresponding dip in accuracy soon after the beginning of the test period, suggesting some nonstationarity in the data or the state of the detector.
This is not surprising, especially since the model was trained on only about three hours of data; it should be taken into account that this method would likely become increasingly susceptible to such issues the less training data it sees and the more time has passed since training.

To illustrate one way this method might be used in practice, we also performed an experiment in which we recorded our model's estimate of glitch probability over a continuous segment of time and compared the output to the actual locations of glitches during that segment.
For this experiment we used 64 seconds of data during \gls{o3} beginning at GPS time 1,238,900,480.
The results of this experiment are shown in Fig. \ref{fig:lr_prob_over_time}, illustrating that the classifier's output largely does correctly indicate the presence or absence of a glitch over time.
If the trained classifier were fed the auxiliary channel data continuously in this fashion, it could potentially be used in near--real time to corroborate the likelihood that a potential event appearing in the strain at any given time is astrophysical or instrumental in origin, independently of other existing two-class glitch classification systems used as automatic, low-latency response to candidate events \cite{essick_thesis}.
By considering more auxiliary channels and decoupling glitch identification from the \gls{gw} strain, our EMU method provides important independent verification of glitches as well as more information about the detector state.

\subsection{Astrophysical implications}\label{sec:astrophysical_implications}
We illustrate the implications of these results in an astrophysical setting with the following examples:

In a use-case scenario where we aim to recover so-called \textit{subthreshold} events that are \textit{lost} by traditional \gls{gw} search methods because they are not classified as detections, it is admissible to have a sizeable false dismissal rate of real signals, but it is desired to have a high glitch rejection rate.
We find that typically a $\sim$65\% glitch rejection rate (i.e. a 35\% false negative rate) in individual detectors can be assumed at the cost of $\lesssim$0.3\% false positives for the high-\gls{snr} case (see, e.g., high-\gls{snr} \gls{o3} curve in Fig. \ref{fig:roc_ER14_O3}).
The proposed method considers data from individual detectors independently, so with such a threshold the chance of a coincident false negative at all three sites is less than 5\% and the chance of a false positive at one or more sites is $\sim$1\%.
Therefore a $\sim$95\% reduction in triple detector coincident glitches corresponds to a negligible chance (i.e., $\sim$O$(1\%)$) to miss a true \gls{gw} signal (i.e., the false positive rate of the glitch rejection is small).

Since in this scenario it is sufficient to flag any of the glitches contributing to the triple coincidence, the method can lead to approximately an order of magnitude reduction in glitch-dominated \gls{far} for triple-detector events.
Let us consider a fiducial sub-threshold \gls{far} of $10^{-7}$~Hz for a transient event candidate that is not sufficient for detection claims, and assume that the \gls{far} is only determined by triple detector glitch rate from these glitches.
If we decrease the triple-detector glitch rate by over an order of magnitude, then the hypothetical \gls{gw} event candidate events moves to the detectable \gls{far} region of $10^{-8}$~Hz.

In an another use case scenario, we can consider all transients detected.
It is then imperative to have a very low false dismissal possibility for real signal, so we can choose a different strategy and operate at a different set point on the \gls{roc} curves displayed on Fig.~\ref{fig:roc_ER14_O3}.
For example, considering that current observation runs produce $\sim$O$(100)$ discoveries, one might require that the false dismissal probability to accidentally miss a true \gls{gw} signal be less than $\sim$O$(0.1\%)$.
We can then require that all three detectors' glitch is flagged to have a triple coincident glitch flagged.
Consequently we need to operate at or below the $\sim$O$(10\%)$ false dismissal rate.
This corresponds to $\sim$90\% individual true positive rate on the ROC curves displayed on Fig.~\ref{fig:roc_ER14_O3}, resulting in a factor of several reduction in the triple-detector glitch rate.

These examples indicate some of the astrophysical opportunities presented by the results here.

\subsection{Glitch subsets}\label{sec:glitch_subsets}
For all of the results presented previously, we considered all triggers recovered by Omicron together without regard to any of their parameters (except for the high-\gls{snr} subsets illustrated in Fig. \ref{fig:roc_ER14_O3}). 
However, the EMU method could also be used independently on subsets of glitches, which could be defined according to any desired criteria, such as frequency, duration, or other Omicron parameters; suspected origin; or by using any of the existing methods that attempt to identify groups of related glitches \cite{2017CQGra..34f4003Z}.
One would expect that some groups of glitches might be easier to classify than others; for example, we noticed that glitches with higher peak frequency or longer duration were generally easier to classify than those with lower peak frequency or shorter duration.

To illustrate this, we performed an experiment in which, prior to training our model, we performed $k$-means clustering \cite{Bishop, Hastie} on the duration, peak frequency, bandwidth, and \gls{snr} (each as reported by Omicron) of glitches in our \gls{er14} training dataset.
$k$-means is a standard clustering algorithm that attempts to identify clusters of related points in a provided dataset.
It outputs the centroid of each cluster and the assignment of each data point to a cluster.

\begin{figure}
    \centering
    \includegraphics[width=\linewidth]{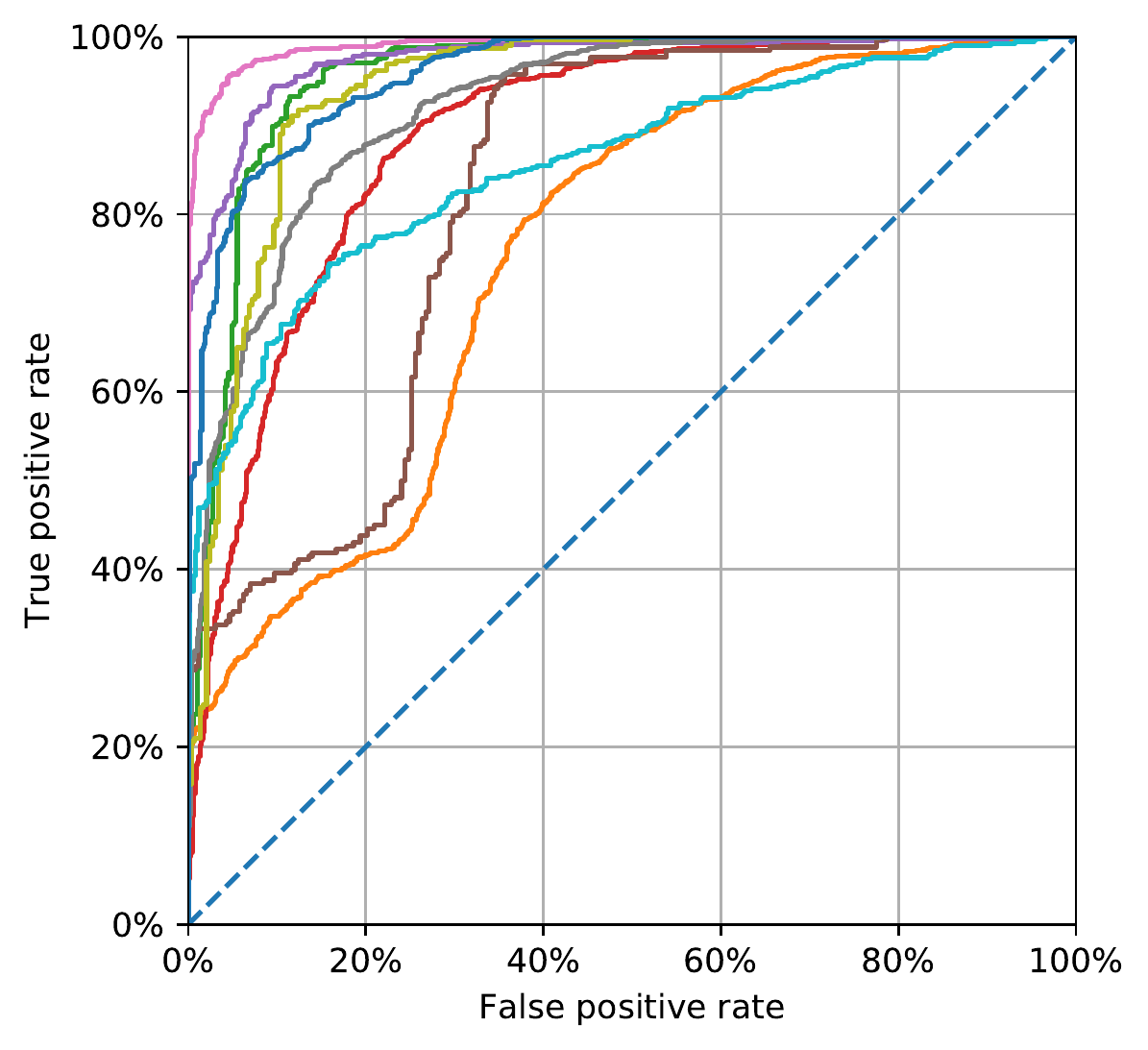}
    \caption{ROC curves for ten subsets of glitches from the \gls{er14} dataset, pre-grouped by $k$-means clustering on their Omicron parameters. We find best performance (the pink curve) on a cluster of high duration, bandwidth, signal-to-noise ratio, and peak frequency.
    \label{fig:ROC_clust}}
\end{figure}

We divided the \gls{er14} training dataset into 10 subsets (10 was chosen arbitrarily) identified by $k$-means and trained 10 elastic net logistic regression classifiers, using one of the glitch subsets as the positive class and all glitch-free samples as the negative class for each classifier.
For testing, we then used the cluster centroids computed on the training set to assign each data point of the \gls{er14} validation dataset to one of the 10 classes and evaluated the performance of each classifier on its corresponding validation subset.
The results are shown in Fig. \ref{fig:ROC_clust}.

We note that the best performing cluster corresponded to glitches with relatively high bandwidth, signal-to-noise ratio, and peak frequency, and relatively long duration.
While several works have attempted to identify auxiliary channels correlated with groups of glitches determined by various means \cite{2017CQGra..34f4003Z,2011CQGra..28w5005S}, none of them has considered the full set of auxiliary channels; our EMU method would enable them to do so, agnostic to the grouping method used.
We leave further exploration of this method and phenomenon to future work.

\section{Conclusions}\label{sec:conclusion}
We have presented a method for flagging potentially glitchy segments of LIGO data using only auxiliary channel data.
It is the first method, to our knowledge, that considers \textit{all} auxiliary channels, and unlike existing methods it does not directly consider the \gls{gw} strain.
As such, it can provide independent corroboration that an event of interest is likely to be astrophysical in origin or not.
It uses a well-established, easily interpretable, and efficient machine learning method.

We report a typical overall accuracy of approximately 80\%, tested on segments from \gls{llo} during \gls{er14} and \gls{o3}, and find that the performance can be improved for certain subsets of glitches.
We also show that this method is capable of reducing glitch-related false detections with negligible false dismissal for a detector network.
The method also provides interpretable results, indicating specific auxiliary channel behavior associated with glitches and its predictions, which could facilitate detector improvements.

We note that the performance characteristics of the method are adjustable and can be modified to suit the requirements of a given study.
Because the method outputs a probability estimate on a continuous scale for each time sample, one can vary the threshold used to classify a sample as glitchy or clean as necessary to achieve a desired balance between false-positive and false-negative rates.

The method can also be trained on a subset of glitches chosen based on characteristics such as \gls{snr}, frequency, or time duration---or any other desired criteria---to target its performance towards similar glitches.
We can also specify a subset of the channels to consider in the feature set.
This can be used to focus on a specific detector subsystem or specific application.
We leave the investigation of these applications to future work.
We also note that the goal of identifying glitches using auxiliary channels is method-independent, and other algorithms can be tested and compared to these results.

The classification model considers a set of features derived from the raw auxiliary channel data at and around a given time.
The feature set was chosen in an ad-hoc manner based on intuition, and we did not significantly attempt to engineer it for performance.
We believe the simplicity of the features is an advantage considering the size of our dataset and illustrates the robustness of the machine learning model we employ.
We note, however, that the model is agnostic to the features used---exactly the same type of model could be employed on top of more advanced, better optimized features.
Considering the engineering of features more carefully represents a worthwhile direction for future work.

Finally, the method presented is not limited to \gls{gw} detectors, and represents a general approach to analysis of the status of a complex system using large numbers of features as input.
It could also be employed in studying other complex machines, experiments, or applications involving similarly high-dimensional data.


\begin{acknowledgments}
We acknowledge computing resources from Columbia University's Shared Research Computing Facility project, which is supported by NIH Research Facility Improvement Grant 1G20RR030893-01, and associated funds from the New York State Empire State Development, Division of Science Technology and Innovation (NYSTAR) Contract C090171, both awarded April 15, 2010.

The authors are grateful for the LIGO Scientific Collaboration review of the paper and this paper is assigned a LIGO DCC number(P1900303). The authors acknowledge the LIGO Collaboration for the production of data used in this study and the LIGO Laboratory for enabling Omicron trigger generation on its computing resources (National Science Foundation Grants PHY-0757058 and PHY-0823459). The authors are grateful to the authors and maintainers of the Omicron and Omega pipelines, the LIGO Commissioning and Detector Characterization Teams and LSC expert Colleagues whose fundamental work on the LIGO detectors enabled the data used in this paper. The authors would like to thank Stefan Countryman and William Tse for their help and suggestions, as well as colleagues of the LIGO Scientific Collaboration and the Virgo Collaboration for their help and useful comments, in particular Joe Betzweiser, Marco Cavaglia, Sheila Dwyer, Reed Essick, Patrick Godwin, and Jess McIver which we hereby gratefully acknowledge.

The authors thank the University of Florida and Columbia University in the City of New York for their generous support.
The authors are grateful for the generous support of the National Science Foundation under grant CCF-1740391.

\end{acknowledgments}

\bibliographystyle{apsrev4-1}
\bibliography{LIGO_glitch_ML}

\end{document}